\documentclass{iopart}

\usepackage{color}
\usepackage[sort&compress]{natbib}
\def\eqref #1{(\ref{#1})}

\usepackage{amssymb}
\usepackage{hyperref}
\usepackage{graphicx}

\newcommand{\resp}{{\cal R}}

\newcommand{\figr}{figure\ }
\newcommand{\Figr}{Figure\ }
\newcommand{\tabl}{table\ }

\newcommand{\eqx}{}
\newcommand{\eqxs}{}

\begin{document}

\title[Price impact without order book]{Price impact without order book: A study of the OTC credit index market}
\author{Z Eisler$^1$ and J-P Bouchaud$^1$}
\address{$^1$ Capital Fund Management, 23 rue de l'Universit\'e, 75007 Paris}
\ead{zoltan.eisler@cfm.fr}
\date{\today}

\begin{abstract}
We present a study of price impact in the over-the-counter credit index market, where no limit order book is used. Contracts are traded via dealers, that compete for the orders of clients. Despite this distinct microstructure, we successfully apply the propagator technique to estimate the price impact of individual transactions. Because orders are typically split less than in multilateral markets, impact is observed to be mainly permanent, in line with theoretical expectations. A simple method is presented to correct for errors in our classification of trades between buying and selling. We find a very significant, temporary increase in order flow correlations during late 2015 and early 2016, which we attribute to increased order splitting or herding among investors. We also find indications that orders advertised to less dealers may have lower price impact. Quantitative results are compatible with earlier findings in other more classical markets, further supporting the argument that price impact is a universal phenomenon, to a large degree independent of market microstructure.
\end{abstract}

\maketitle

\section{Introduction}
\label{sec:introduction}

\subsection{Price impact}

Liquidity in financial markets is an elusive concept, with many definitions in existence. From a practical point of view, however, one of its most important metrics is the response of price to buying and selling.
This reaction is called \emph{price impact}, and it has been treated in a long series of empirical papers (see e.g. \citet{hasbrouck1991measuring,bouchaud2004fluctuations,almgren2005direct, moro2009market, bouchaud2008markets,toth2011anomalous} and refs. therein). One of the most important findings is that impact is not only mechanical but dynamic, meaning that it cannot be described exclusively by the revealed supply or demand at any given time -- 
say the content of the visible limit order book \citep{weber2005order}. It is rather related to underlying ``latent'' supply and demand \citep{toth2011anomalous,donier2015fully} which correspond to the intentions of market participants, and which manifest themselves over time. Most of the recent studies of impact have been carried out on transparent, listed markets. Nevertheless, many aspects of price impact appear to be universal, i.e. common to many asset classes, even to exotic ones \citep{donier2015million, toth2016square}.

In this paper, we will continue the exploration by presenting the price impact of individual transactions in credit indices, where trading does not take place in limit order books. The remainder of this section introduces these products, briefly surveys the relevant literature, and presents the data used for our study. Section \ref{sec:naive} then describes a naive approach to calculate impact based on a standard propagator technique. 
Section \ref{sec:correction} looks at the effect of misclassification between buy and sell trades, and corrects the resulting biases in our results. Section \ref{sec:seasonal} discusses a temporary pattern of increased order splitting and higher impact observed in the data. Section \ref{sec:on_off_sef} finds indications that the impact of trades increases with the number of dealers involved. Finally Section \ref{sec:conclusion} concludes.

\subsection{The credit index market}

Today the credit index market is fairly mature. The most liquid derivative products are proposed by Markit, we will look at four of them: two US based indices CDX IG (for Investment Grade) and CDX HY (for High Yield), and their European counterparts iTraxx Europe and iTraxx Crossover, respectively. These correspond to baskets of CDSs \citep{oehmke2014anatomy}, each of which represents an insurance on bonds of a given corporate issuer within the respective grade and geographical zone. The instruments are standardized, their mechanics very much resembles futures contracts, and they roll once every 6 months. At the time of the roll a new maturity is issued, these are called ``series", for example S23, S24 and S25 for iTraxx Europe. On any given day most of the liquidity is concentrated in the most recently issued five-year series of each index at the time, in the following we will only study these (see \tabl \ref{tab:products}).

One particularity of this market -- as opposed to stocks or futures -- is that it is purely \emph{over-the-counter} (OTC), currently \emph{without} any liquid limit order book. Information is fragmented, there is no single, central source to verify when one is looking for tradable prices, client orders pass through a large number of dealers instead. The latter usually do provide indicative bid/ask prices, but the actual trades are mostly done via Request for Quotes (RFQ), see also \citet{hendershott2015click}. This means that the client (liquidity taker) auctions off its trades to the liquidity providers, by sending them information about the conditions of the deal (which product, buy or sell, size) either electronically or by voice, receiving competing quotes in return, and taking the best price.

To counteract the bilateral design of OTC markets which favors opacity, the Dodd-Frank Act has mandated several changes. Among them are obligatory post-trade reporting, and the creation of Swap Exchange Facilities (SEFs) which provide an organized framework to dealing in eligible OTC instruments. Today a large portion of trades is required to go through SEFs, whose volume is predominantly done via electronic RFQ. Even though they provide order books, those are -- for the moment -- empty.

\subsection{Literature review}

Credit trading has received considerably less attention than equities or futures, and most studies have been done by or in collaboration with regulators who have privileged access to non-anonymous data. \citet{gehde2015liquidity} study records from the German Bundesbank regarding single-name CDS issues. They find significant price impact using a model where the effect of each trade is permanent. \citet{shachar2012exposing} of the New York Federal Reserve defines buy/sell orders by assuming that the initiator of the trades is the end-user (as opposed to the dealer), and focuses to a large extent on the inventory management of dealers. The study finds evidence of "hot-potato" trading \citep{lyons1997hotpotato} whereby an initial client trade changes hands among dealers several times, while its effect is being gradually incorporated into the price. \citet{loon2016does} of the Securities and Exchange Commission focus on the same credit indices as our study. Their work takes a policy-maker's point of view, and argues that the wider transparency created by the Dodd-Frank Act has improved several metrics of liquidity. They focus particularly on the transitory period during the introduction of the reform, whereas we will consider more recent data where market structure is already relatively stable. Finally, \citet{hendershott2015click} analyze both electronic and voice trading in single-name CDS. Most notably they identify price impact related to information leakage, especially when the client requests prices from many dealers, and even if he finally decides not to trade.

\subsection{The dataset}

Our period of study is 17 June 2015 -- 31 August 2016. For the four products we have recorded a semi-realtime (several updates per minute) indicative data feed via a service called CBBT (Composite Bloomberg Bond Trader). This represents a continuous, electronic poll of recent executable prices from dealers. Nevertheless, it is not a bid or ask price, only an indicative level around which one expects to be able to transact. At some time $t$ the indicative price is quoted as a credit spread $s_t$, which is the annualized insurance premium in basis points.\footnote{Note that the quoting convention for CDX HY is different from the rest, but the credit spread can be recalculated based on the available prices.} In the following we will express all prices as basis points of the typical credit spread itself, meaning
\begin{equation*}
m_t = 10^4 \times \frac{s_t}{\left\langle s_t\right\rangle},
\end{equation*}
where $\left\langle \cdot\right\rangle$ denotes a time average.

Anonymous information about trades is also available from a different source: \emph{trade repositories} mandated by regulation. We have used the records of two such organizations.\footnote{Data from Bloomberg SDRV is available at \href{http://www.bloombergsdr.com/}{http://www.bloombergsdr.com/}, and from the DTCC at \href{http://www.dtcc.com/repository-otc-data}{http://www.dtcc.com/repository-otc-data}.} These include a substantial part of all trades with credit spread, timestamp, volume and other additional information.

While the data are rich and relatively clean, the two sources (prices and trades) are independent, and there is no \emph{a priori} reason for perfect synchronization between the two.

\begin{table}
\caption{Summary of the different credit index products studied. Notice that the value of $G$ is much more stable than that of $\resp$ across different series of the same product.}
\begin{indented}
\item[]\begin{tabular}{@{}lcccc}
\br
\multicolumn{1}{c}{Product code} & Avg. intertrade & $N_\mathrm{eff}$ & ${\cal R}(\ell=30)$ & $G(\ell=30)$ \\
& time [sec] & & [bps] & [bps] \\
\mr
CDX HY S24           &                         75 &              2.3 &               5.8 &                      3.4 \\
CDX HY S25           &                         66 &             14.2 &              51.2 &                      3.9 \\
CDX HY S26           &                         90 &              1.7 &               7.1 &                      4.2 \\ \hline
CDX IG S24           &                        113 &              1.6 &              11.8 &                      7.5 \\
CDX IG S25           &                         92 &              9.8 &              55.0 &                      4.9 \\
CDX IG S26           &                        119 &              2.8 &              10.0 &                      5.3 \\ \hline
iTraxx Crossover S23 &                        128 &              1.5 &              13.4 &                      9.3 \\ 
iTraxx Crossover S24 &                        106 &              5.7 &              65.4 &                      6.6 \\
iTraxx Crossover S25 &                        199 &              1.2 &              19.9 &                     10.6 \\ \hline
iTraxx Europe S23    &                        171 &              1.4 &              15.8 &                     10.1 \\
iTraxx Europe S24    &                        132 &              7.1 &              67.1 &                      8.4 \\
iTraxx Europe S25    &                        181 &              2.0 &              21.4 &                     10.2 \\
\br
\end{tabular}
\label{tab:products}
\end{indented}
\end{table}

\begin{figure}[tbh]
\begin{center}
\includegraphics[width=0.9\columnwidth]{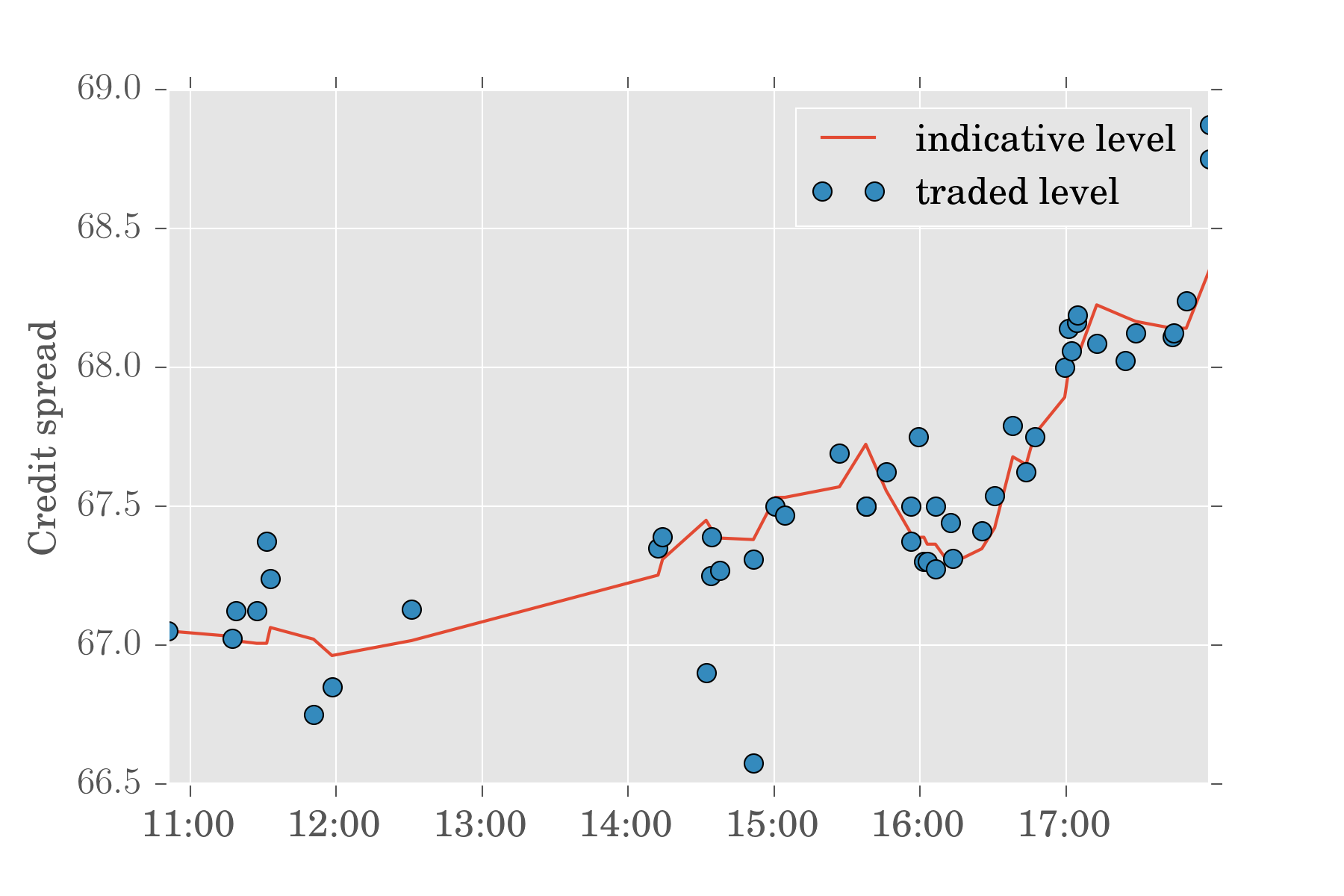}
\caption{An example of the time evolution of indicative credit spread and its value reported for trades. We show the product iTraxx Europe S25 on 31 August 2016.}
\label{fig:price_example}
\end{center}
\end{figure}

\section{Naive propagators}
\label{sec:naive}

Price impact is often analyzed in the context of linear models, where the market price $m_t$ just before trade $t$ is written as a linear combination of the time dependent impact of past trades \citep{bouchaud2004fluctuations}:
\begin{equation}
    m_t = \sum_{t'<t}\left[{G}(t-t')\epsilon_{t'} + \eta_{t'} \right] + m_{-\infty}.
    \label{eq:ptMO}
\end{equation}
$\epsilon_{t'}$ is the sign of the trade at time $t'$ ($+$ for buyer, $-$ for seller initiated trades), and $\eta_{t'}$ is an independent noise term. ${G}(\ell)$ is called the `propagator', and it describes how the price at time $t$ is modified \emph{due to} the trade at $t-\ell$. In equity and futures markets, this propagator is found to decay with time, i.e. a large part of price impact is transient rather than permanent (for a recent study of the long term behaviour of $G(\ell)$ in equities, see \citet{brokmann2015slow}).

In order to calibrate the model \eqref{eq:ptMO} one calculates the \emph{response function} ${\cal R}(\ell)$, which is defined as
\begin{equation}
    {\cal R}(\ell) = \langle (m_{t+\ell}-m_t) \cdot \epsilon_t \rangle,
\end{equation}
and which quantifies the price move \emph{after} a trade, but not necessarily \emph{due to} the trade. One then measures the autocorrelation of order signs, which is is customarily defined as
\begin{equation}
  C(\ell) = \langle \epsilon_t \epsilon_{t+\ell} \rangle.
\end{equation}
And finally one solves the linear equation \citep{bouchaud2006random}
\begin{equation}
\label{eq:RCG}
    {\cal R}(\ell)  = \sum_{0 < n\leq \ell} {G}(n) C(\ell-n) + \sum_{n>0} [{G}(n+\ell)-G(n)] C(n)
\end{equation}
to map out the numerical value of $G(\ell)$.\footnote{In fact, it is much better to solve numerically the corresponding equation for the discrete derivative of $G$ in terms of the discrete derivative of $R$. This is what we have done in this paper.}

To guess the sign of trades one often relies on some heuristic. If we denote the price of trade $t$ by $p_t$, then simply
\begin{equation}
  \label{eq:sign}
  \epsilon_t = \mathrm{sign}(p_t-m_t).
\end{equation}
An example of transaction and reference prices is shown in \figr \ref{fig:price_example}.

\begin{figure}[tbh]
\begin{center}
\includegraphics[width=1.0\columnwidth]{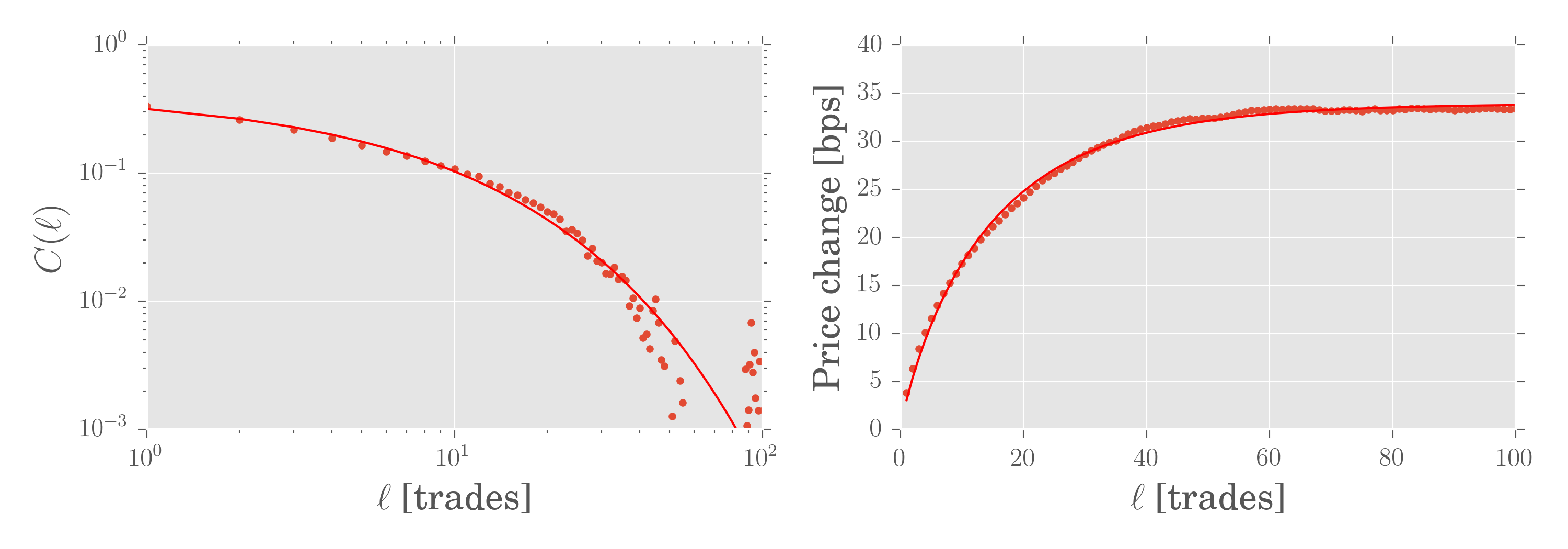} 
\caption{{\it (left)} Average of the sign autocorrelation $C(\ell)$, and the stretched exponential fit $a\times \exp(-b\ell)^{\nu}$, with $a=0.43$, $b=0.16$ and $\nu=2/3$. {\it (right)} Average of the response function $\resp(\ell)$, and the stretched exponential fit $a\times [1-\exp(-b\ell)^{\nu}]$, with $a=33.9$, $b=0.068$ and $\nu=0.88$.}
\label{fig:C}
\end{center}
\end{figure}

As one can see from \figr \ref{fig:C}, the shape of $C(\ell)$ is well fitted by a stretched exponential. This is true for most individual products and on average across them, and it is in contrast with earlier studies in order book markets, where $C(\ell)$ rather decays as a slow, power-law function \citep{bouchaud2008markets}. In the latter liquidity at good prices is often small \citep{bouchaud2006random}, so large orders tend to be sliced, producing a long-range autocorrelation of small trades. In OTC markets clients are encouraged to request deals that correspond to their full liquidity needs, so that after the trade is done, the dealer offloading the inventory just acquired will not have to compete for liquidity with the same client. Since trades are bilateral, the dealer knows the identity of the client, and can reward or penalize it by adjusting the bid-ask spread according to any adverse selection perceived on earlier deals \citep{osler2016bid}.

Since the order sign process is not long range correlated, its autocorrelation function is integrable and one can therefore define an \emph{effective number of correlated orders} via
\begin{equation*}
N_\mathrm{eff} = \sum_{\ell = 0}^\infty C(\ell).
\end{equation*}

This value varies between $1.2$ and $14.2$, as reported in \tabl~\ref{tab:products}. These differences are due to time periods, we will study this point further in Section \ref{sec:seasonal}

\begin{figure}[tbh]
\begin{center}
\includegraphics[width=0.9\columnwidth]{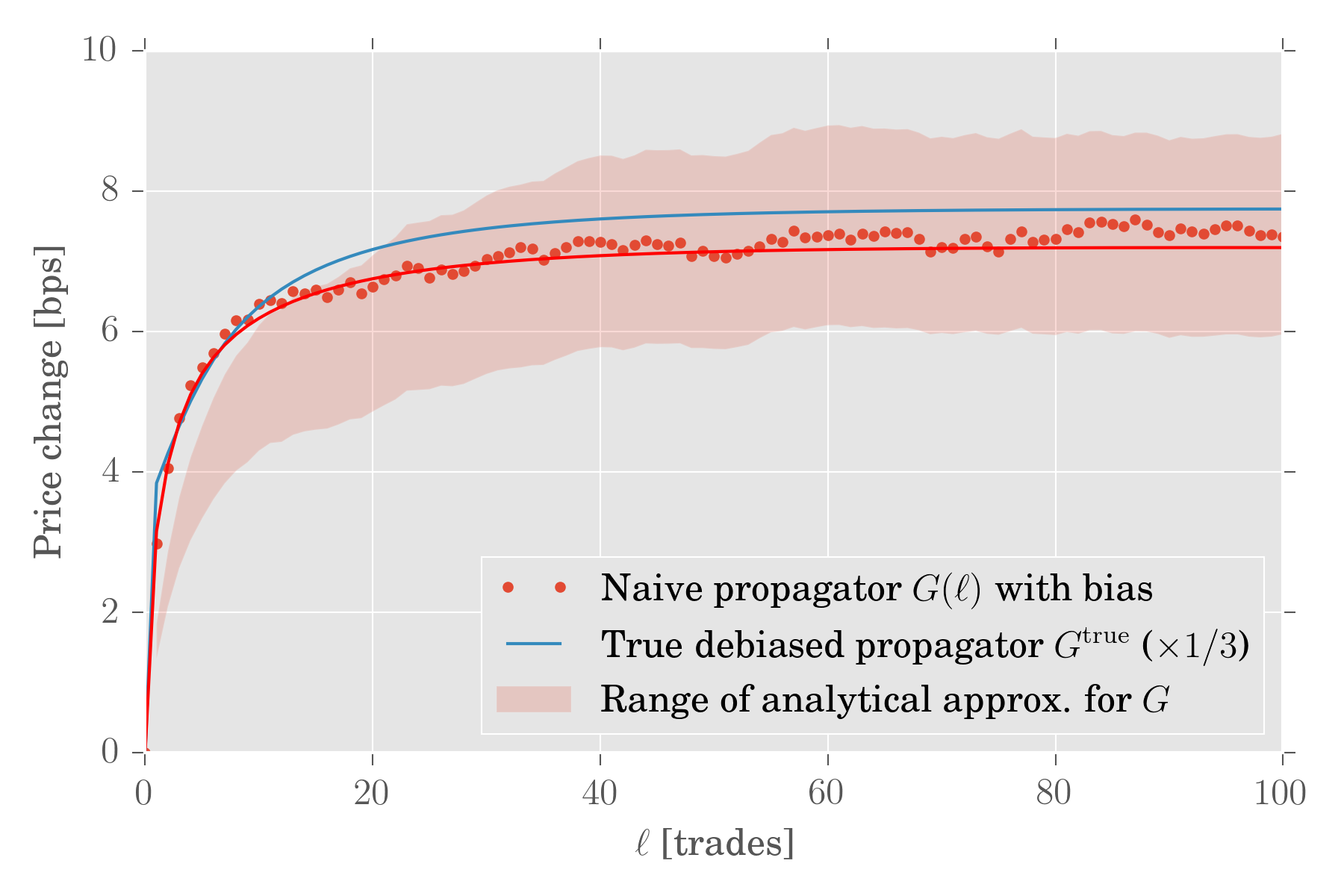} 
\caption{Average value across products of various quantities: ({\it red points}) Propagator $G(\ell)$, ({\it red line}) fits of the former quantity with the same stretched exponential form as above. ({\it red shaded area}) Bounds on $G(\ell)$ based on \eqxs \eqref{eq:resplower} and \eqref{eq:respupper}. Note that as expected these are only valid for large $\ell$.  ({\it blue line}) The true, bias-adjusted propagator $G^\mathrm{true}(\ell)$ calculated with the fits of $C^\mathrm{true}(\ell)$ and $\resp^\mathrm{true}(\ell)$, and divided by a factor $3$ for better readability.}
\label{fig:Gtrue}
\end{center}
\end{figure}

We can calculate the propagator via \eqx \eqref{eq:RCG}, their average across products is given in \figr \ref{fig:Gtrue}. \citet{bouchaud2004fluctuations} have shown that if $C(\ell)$ has a power-law form, then the propagator should itself decay over time as a power-law in order to maintain the efficiency of prices. In other words, impact is mostly {\it transient} in that case. On the other hand, since our $C(\ell)$ is short ranged, the same argument predicts that $G(\ell)$ should tend to a constant for large $\ell$, corresponding to a non-zero {\it permanent impact} component. This is indeed what we observe, see \figr \ref{fig:Gtrue}.

Calculating $G$ from $\resp$ involves inverting a matrix whose elements are related to $C$. This operation amplifies the noise in the correlations, and so it is useful to also give approximate formulas that avoid this. If we know that $G$ is increasing before converging to a fixed value, and $C$ is positive, then from \eqx \eqref{eq:RCG} one can find two bounds on $G(\ell)$. In the limit when $C(\ell)$ reaches zero much more quickly than $G(\ell)$ goes to its asymptotic value, we can write for large $\ell$ that
\begin{equation}
(2 N_\mathrm{eff}-1)^{-1}\resp(\ell) \leq G(\ell).
\label{eq:resplower}
\end{equation}
Conversely, if we assume that $G$ saturates immediately, then $\resp$ will be maximal, and for large $\ell$ we get
\begin{equation}
G(\ell)\leq N_\mathrm{eff}^{-1} \resp(\ell).
\label{eq:respupper}
\end{equation}
\Figr \ref{fig:Gtrue} shows these bounds which are not excessively wide, as well as the real $G$ calculated numerically.

In reality the propagator -- which was not observed in previous studies -- increases steeply but continuously in the initial period. This makes sense in the absence of a central orderbook: The information that someone bought or sold takes a finite time of $5-10$ trades to diffuse in the market and to get incorporated into the price.

\begin{figure}[tbh]
\begin{center}
\includegraphics[width=0.9\columnwidth]{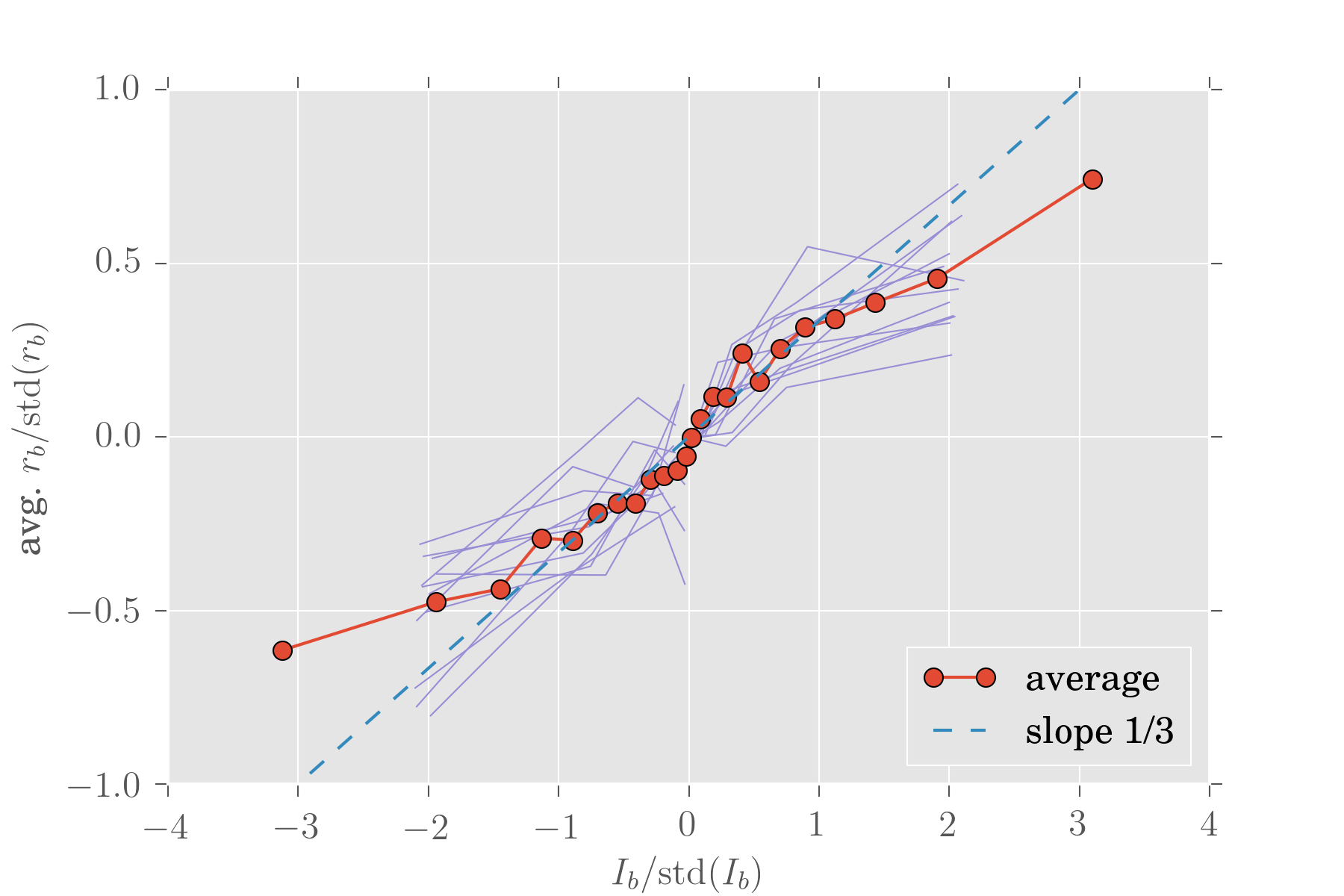} 
\caption{The average return in 15-minute windows normalized by their absolute mean ($r_b/\left\langle|r_b|\right\rangle$), as a function of the imbalance normalized by its own standard deviation ($I_b/\mathrm{std}(I_b)$). Data points have been separated into 30 groups according to their rank on the horizontal axis, we show the average returns in each group. Outliers outside the horizontal range $[-3, +3]$ have been discarded. The purple lines correspond to individual products, and the red points to all products together. The dashed line is a linear fit with slope $1/3$. Note the clear concavity of the average curve as the volume imbalance increases.}
\label{fig:bins15}
\end{center}
\end{figure}

Beyond the propagators which give a microscopic description of price moves, one can also look at a more aggregate characterization by dividing the data into $15$ minute bins. Then one can calculate in each bin $b$ the net signed notional defined as
\begin{equation*}
I_b = \sum_{t\in b} \epsilon_tQ_t,
\end{equation*}
where the sum runs over trades in the bin, and $Q_t$ is the notional value of trade $t$. As a function of this quantity one can calculate the average price change $r_b = m_b - m_{b-1}$ over the bin. \Figr \ref{fig:bins15} confirms a strong correlation, similar results can be obtained regardless of the precise time scale. Note the concavity of this plot, also observed in many other markets with order books (see e.g. \citep{bouchaud2008markets}, \figr 2.5). Although we do not have enough statistics to test on our trades the $\sqrt{Q}$ impact law universally observed in all markets studied so far, we believe that the concave shape seen in \figr \ref{fig:bins15} is compatible with the general ``latent liquidity'' idea of \citet{toth2011anomalous,donier2015fully}.

\section{Correction for the noise in order signs}
\label{sec:correction}

In the previous section we have confirmed the existence of price impact in OTC credit indices. However, the magnitude of the effect remains to be validated for the following reason. It is known that the heuristic \eqref{eq:sign} for identifying order signs does not always give correct results. If $\epsilon_t$ is incorrect, naturally all expectation values calculated from it will be incorrect as well. In order to verify such biases in our results, we are going to use a proprietary dataset including 252 trades executed over the same period by our firm (CFM).

As opposed to the detected trade sign $\epsilon_t$, let us introduce the notation $\epsilon^\mathrm{true}_t$ for the true sign of the same order, which is not a priori known, except for those of CFM. It is also convenient to introduce an auxiliary variable $q_t$ that is $1$ when we classified the trade correctly, and $0$ when we did not. This way 
\begin{equation}
\epsilon_t = q_t\epsilon^\mathrm{true}_t + (1-q_t)\times(-\epsilon^\mathrm{true}_t) \equiv \epsilon^\mathrm{true}_t (2q_t-1).
\label{eq:q}
\end{equation}
If we look at the above mentioned CFM trades, the rate of correct classification, described by the average $\left\langle q_t\right\rangle_{t\in\mathrm{CFM}}$, is only 72\%. This value is constant within noise level across different months in the sample.

\begin{figure}[tbh]
\begin{center}
\includegraphics[width=0.9\columnwidth]{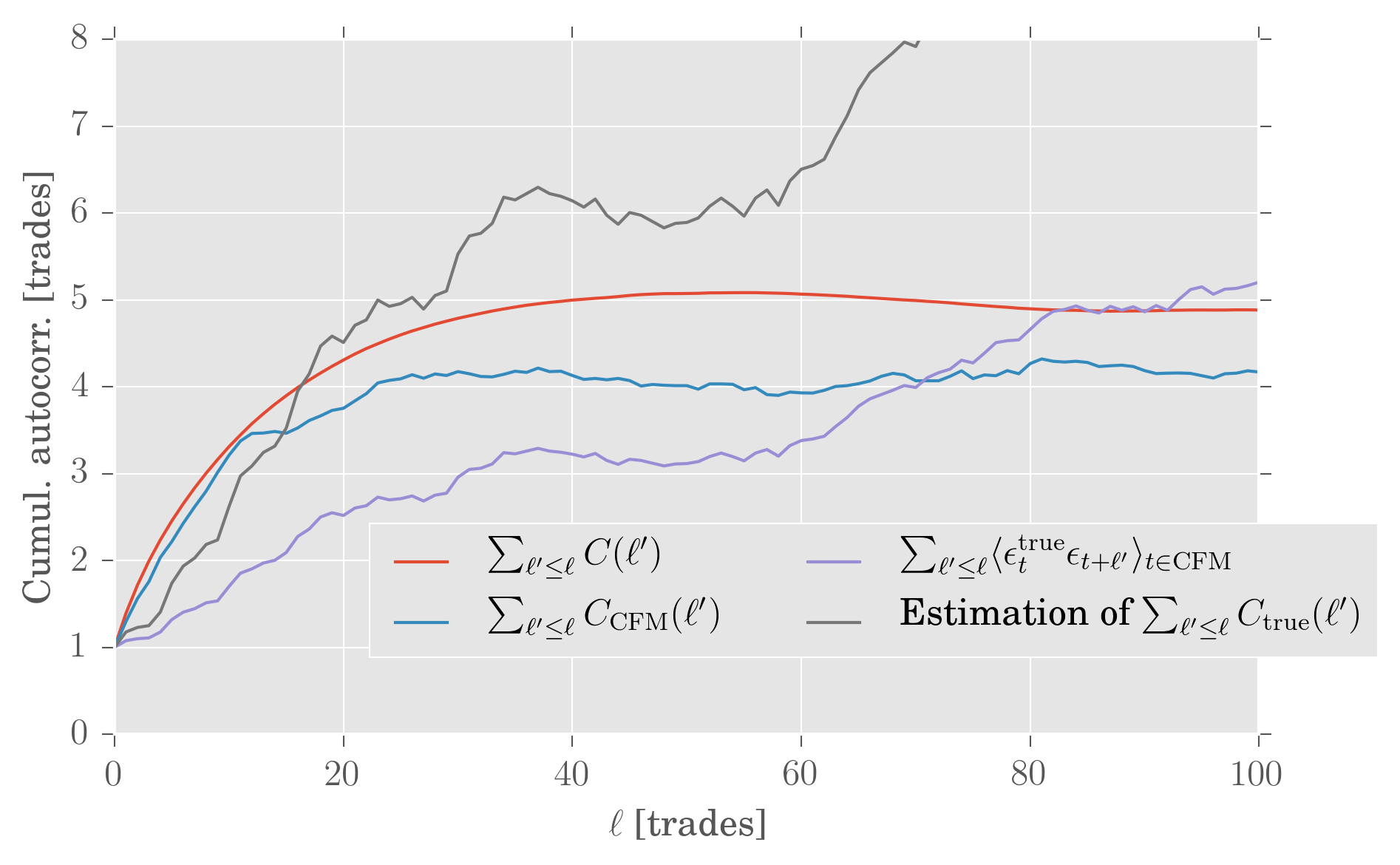} 
\caption{Cumulative sign autocorrelation functions. The estimated curve corresponds to \eqx \eqref{eq:Ctrue}.}
\label{fig:Cint}
\end{center}
\end{figure}

As a first step we would like to show that basic correlations of the detected order signs $\epsilon_t$ are the same in the CFM subset of trades and the rest. Let us compare $C(\ell) = \langle\epsilon_t\epsilon_{t+\ell}\rangle$ with the subsample average 
\begin{equation*}
C_\mathrm{CFM}(\ell) = [\langle\epsilon_t\epsilon_{t+\ell}\rangle_{t\in\mathrm{CFM}}+\langle\epsilon_t\epsilon_{t+\ell}\rangle_{t+\ell\in\mathrm{CFM}}]/2,
\end{equation*}
where we conditioned on at least one of the trades belonging to CFM. \Figr \ref{fig:Cint} shows that there is a fair match, especially for short lags. This gives an indication that other statistics calculated on CFM trades may be approximately similar to the whole market, and can be used in the following.

Let us now look at how misclassification biases our earlier calculations. Let us define
\begin{eqnarray}
C_\mathrm{true}(\ell) = \left \langle \epsilon^\mathrm{true}_t \epsilon^\mathrm{true}_{t+\ell}\right\rangle = \left \langle \epsilon^\mathrm{true}_t \epsilon_{t+\ell}\right\rangle + \left\langle \epsilon^\mathrm{true}_t (\epsilon^\mathrm{true}_{t+\ell}-\epsilon_{t+\ell})\right\rangle.
\end{eqnarray}
We can readily measure the first term when $t\in\mathrm{CFM}$, whereas for $\ell \geq 1$ the second term can be rewritten as
\begin{eqnarray}
  \left\langle\epsilon^\mathrm{true}_t (\epsilon^\mathrm{true}_{t+\ell}-\epsilon_{t+\ell})\right\rangle = -\left\langle\epsilon^\mathrm{true}_t \epsilon^\mathrm{true}_{t+\ell}\cdot 2(q_{t+\ell}-1)\right\rangle \approx \nonumber \\
  \left\langle\epsilon^\mathrm{true}_t\epsilon^\mathrm{true}_{t+\ell} \right\rangle \cdot 2(\left\langle q_{t+\ell}\right\rangle-1) = -2 C^\mathrm{true}(\ell)\cdot (\left\langle q_t\right\rangle-1).
  \label{eq:truedelta}
\end{eqnarray}
For the approximation step we assumed that whether or not we make a mistake in identification is independent of the two-point product of true signs. After reorganization and assuming that we can use CFM trades in part of the correlation, we get
\begin{equation}
  C_\mathrm{true}(\ell \geq 1) = \frac{\left \langle \epsilon^\mathrm{true}_t \epsilon_{t+\ell}\right\rangle_{t\in\mathrm{CFM}}}{2\left\langle q_t\right\rangle-1}.
  \label{eq:Ctrue}
\end{equation}
This estimation of $C_\mathrm{true}(\ell)$ is shown in \figr \ref{fig:Cint}.

As for the response function, one can define its ``true" variant as
\begin{equation}
  {\cal R}^\mathrm{true}(\ell) = \langle (m_{t+\ell}-m_t) \cdot \epsilon^\mathrm{true}_t \rangle.
  \label{eq:Rtrue}
\end{equation}
This is related to the response with detected order signs as
\begin{eqnarray}
  \resp(\ell) = \langle (p_{t+\ell}-p_t)\epsilon_t \rangle = \nonumber \\
  \langle (p_{t+\ell}-p_t)\times(2q_t-1)\epsilon^\mathrm{true}_t\rangle \approx (2\langle q_t\rangle-1)\resp^\mathrm{true}(\ell).
  \label{eq:Rtrue2}
\end{eqnarray}
In the approximation step we neglect the correlation of identification error and future price change. Finally:
\begin{eqnarray}
  \resp^\mathrm{true}(\ell) = \frac{\resp(\ell)}{2\langle q_t\rangle-1}.
  \label{eq:Rtrue3}
\end{eqnarray}

Finally one can define a true propagator $G^\mathrm{true}(\ell)$ via \eqx \eqref{eq:RCG} by inserting ${\cal R}^\mathrm{true}(\ell)$ and $C^\mathrm{true}(\ell)$. If we use \eqxs \eqref{eq:Ctrue} and \eqref{eq:Rtrue3} for the approximation of these latter, one can obtain the numerical value of the true propagator averaged over all products, see \figr \ref{fig:Gtrue}. This shows that finally $G^\mathrm{true}(\ell) \approx 3G(\ell)$.

\section{Seasonal patterns in order splitting}
\label{sec:seasonal}

\begin{figure}[tbh]
\begin{center}
\includegraphics[width=0.9\columnwidth]{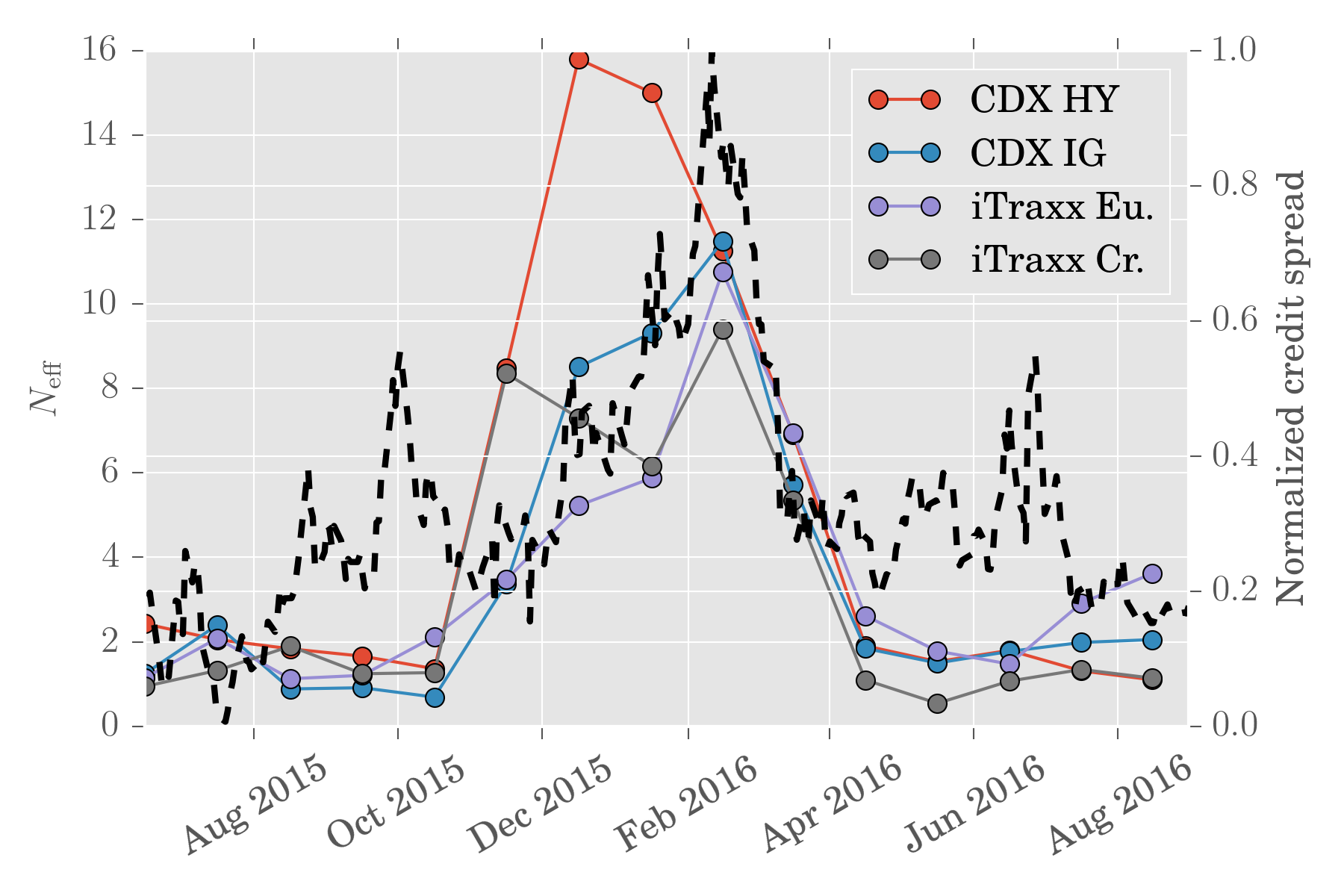} 
\caption{The points in color show the effective number of correlated trades for each product, measured in 1-month periods. The dashed black line represents the credit spread shifted and normalized such that the data spans the interval [0, 1], this is an average over the four products.}
\label{fig:Neff_all}
\end{center}
\end{figure}

\begin{figure}[tbh]
\begin{center}
\includegraphics[width=0.9\columnwidth]{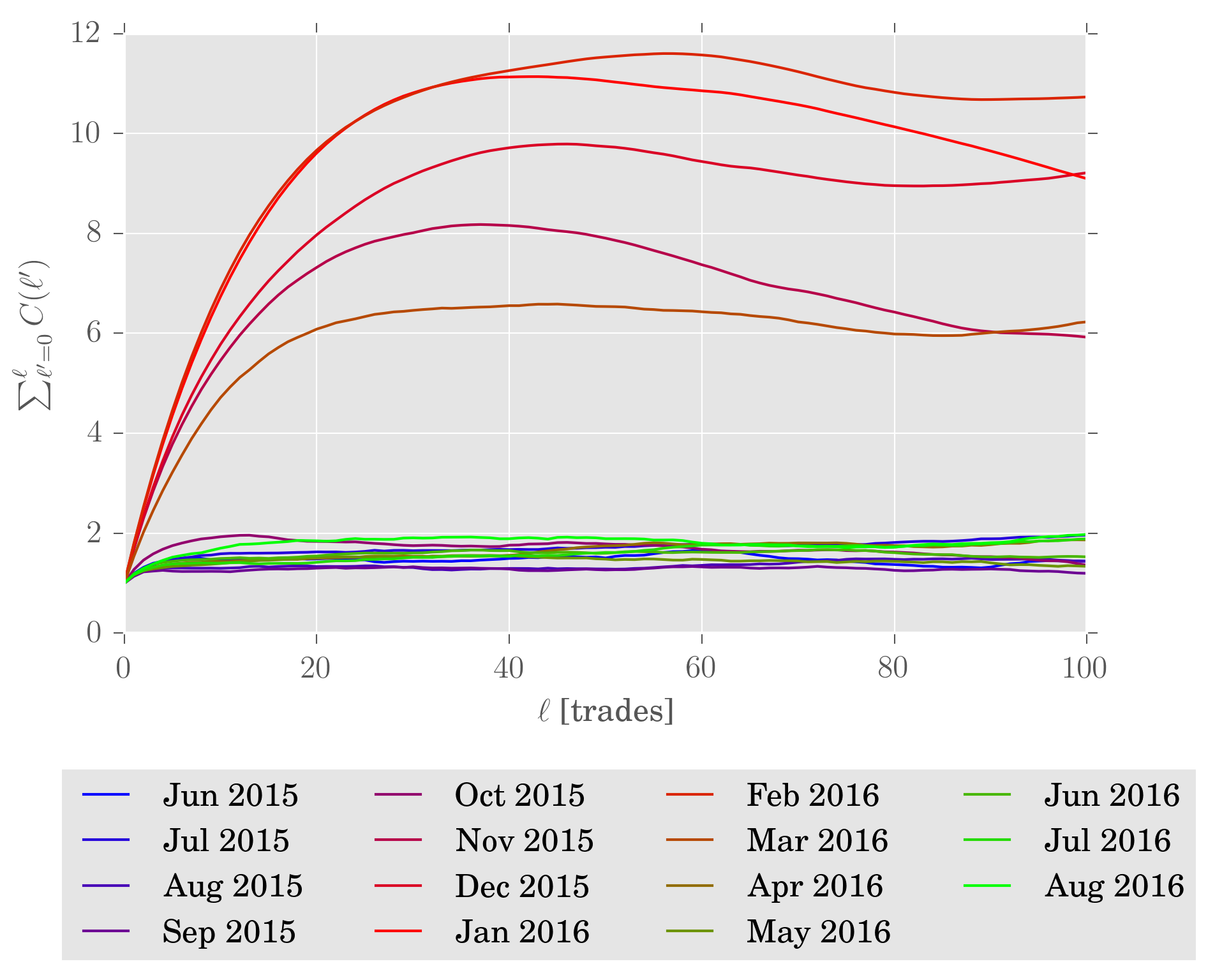} 
\caption{The cumulative trade sign autocorrelation $\sum_{\ell' = 0}^\ell C(\ell')$, averaged across products. Each curve corresponds to a 1-month period.}
\label{fig:Neff_cumsum}
\end{center}
\end{figure}

We now revisit our results to study their variation across time. \Figr \ref{fig:Neff_all} shows the measured value of $N_\mathrm{eff}$ in monthly windows. One can see that order correlations intensify at the turn of the year, while at the same time credit spreads climb to a local maximum. Liquidity itself remains roughly constant. In \figr \ref{fig:Neff_cumsum} we offer a more detailed view of the effect, by showing the cumulative autocorrelation $\sum_{\ell' = 0}^\ell C(\ell')$ for each month separately (averages over all products). We do not see much structure in periods of low correlation, while during high correlation $C$ is very positive up to 10--20 trades. Note that the mis-classifier variable $q_t$ is reasonably stationary, so that this seasonality is too strong to result from an incorrect classification of trades.

\begin{figure}[tbh]
\begin{center}
\includegraphics[width=0.9\columnwidth]{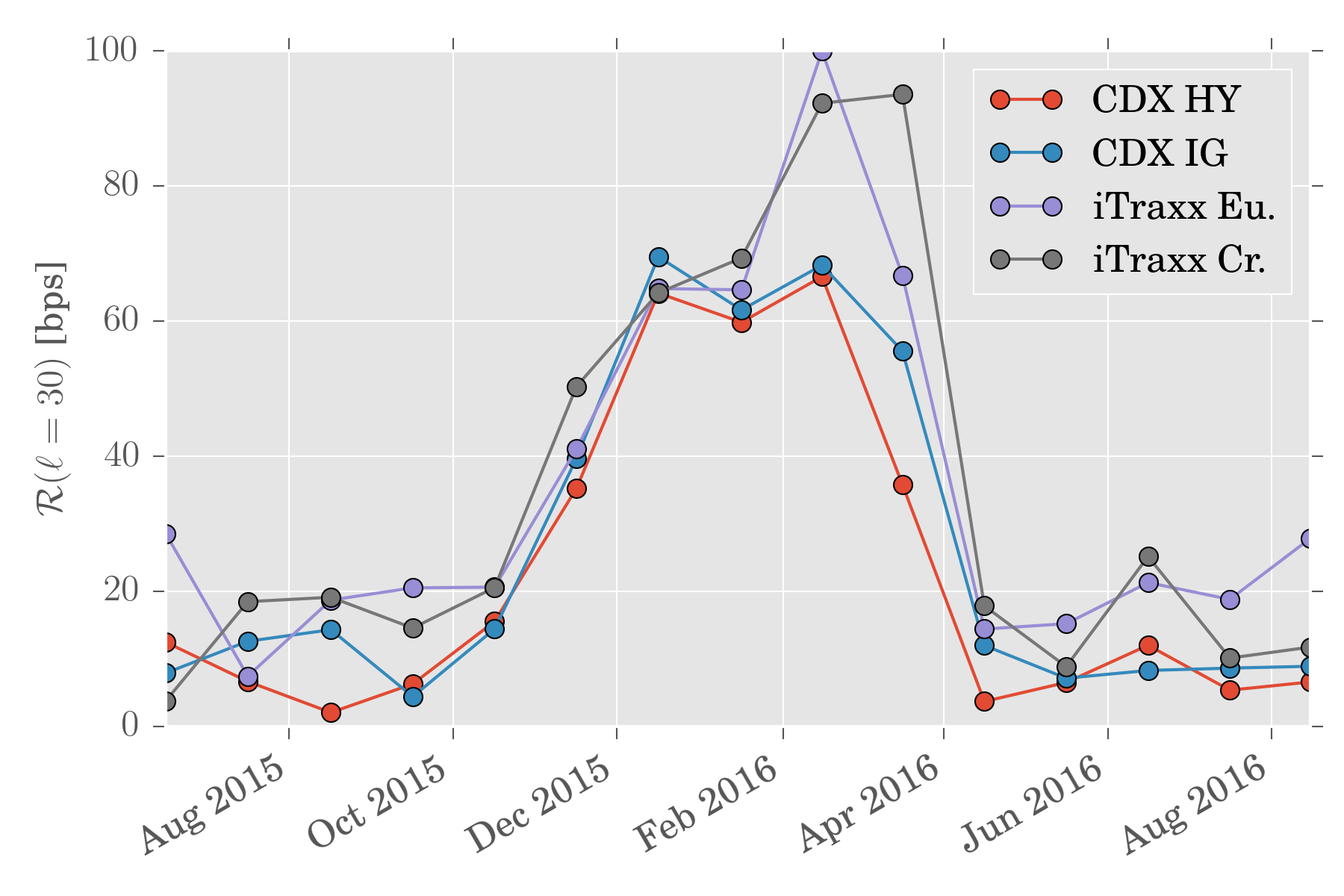} 
\caption{The points in show the response function ${\cal R}$ after 30 trades for the each product, measured in 1-month periods.}
\label{fig:monthly_Reff}
\end{center}
\end{figure}

An explanation could be given in the context of the "hot-potato" theory of \citet{lyons1997hotpotato}, advocated for credit markets in \citet{shachar2012exposing}. Clients are expected to trade the full required size in a single deal, so further orders (and hence $N_\mathrm{eff} > 1$) could come from the subsequent inter-dealer exchange of risk. In the period of difficult markets liquidity becomes "recycled" as it takes a longer time for the position to find a final counter party to warehouse the risk. This, however, does not explain the increase of ${\cal R}$ which is shown in \figr \ref{fig:monthly_Reff}. Response is amplified in proportion to correlations, which means that these additional trades have full impact, and they likely cause a net variation of dealer inventory \citep{shachar2012exposing}. Hence this is more likely the signature of increased real order splitting or herding among clients, in the spirit of recent papers on transparent markets \citep{bouchaud2008markets}.

\section{Comparison of on-SEF and off-SEF trades}
\label{sec:on_off_sef}

\citet{loon2016does} compare various forms of trading, and argue that in the transitory period after the enactment of Dodd-Frank, increased market transparency has lead to lower trading cost and price impact. Their data dates from 2013, when uncleared and off-SEF trading were still commonplace. In our more recent dataset only $11\%$ of all trades are off-SEF, and in terms of transparency we no longer expect much difference. More importantly though, on-SEF it is mandatory to have \emph{at least} three competing brokers when requesting a price, whereas on electronic platforms for off-SEF this is \emph{at most} three brokers. It is common lore among traders that increased competition might reduce instantaneous costs, but as \citet{hendershott2015click} also show, it leads to higher information leakage, and thus more impact.

\begin{figure}[tbh]
\begin{center}
\includegraphics[width=0.9\columnwidth]{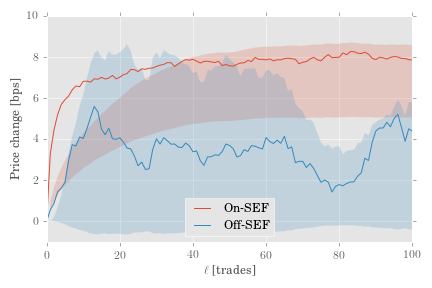}
\caption{Naive propagators for on-SEF and off-SEF trades. The shaded areas correspond to the theoretical bounds derived from the multi-propagator equivalents of \eqref{eq:resplower} and \eqref{eq:respupper}, which are only expected to be valid for large $\ell$.}
\label{fig:on_off_sef}
\end{center}
\end{figure}

\begin{figure}[tbh]
\begin{center}
\includegraphics[width=0.9\columnwidth]{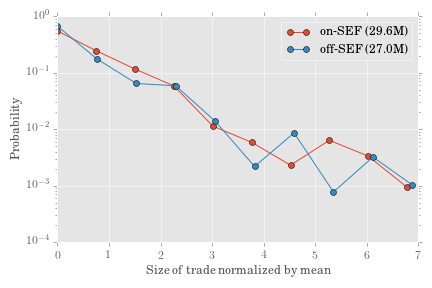}
\caption{The distribution of order sizes on-SEF and off-SEF, close to an exponential. Both are normalized by their (nearly identical) respective means, which are shown in the legend.}
\label{fig:trade_distribution}
\end{center}
\end{figure}

It is straightforward to extend \eqx \eqref{eq:ptMO} to a case where trades are classified into discrete categories $\pi$ \cite{eisler2011models}, each of which has its own propagator $G_\pi$. If each trade $t'$ falls into category $\pi_{t'}$, then
\begin{equation}
    m_t = \sum_{t'<t}\left[G_\mathrm{\pi_{t'}}(t-t')\epsilon_{t'} + \eta_{t'} \right] + m_{-\infty}.
    \label{eq:ptMO2}
\end{equation}
We use this technique to separate on-SEF and off-SEF execution, the naive impact kernels and the corresponding theoretical bounds are shown in \figr \ref{fig:on_off_sef}. Indeed, despite the high noise level we find that off-SEF trades have significantly lower impact than on-SEF ones. This is not a result of the orders themselves being smaller, the mean size and the shape of the distribution are nearly identical, see \figr \ref{fig:trade_distribution}. We see this rather as support for the theory of \citet{hendershott2015click}, that price impact grows with the number of dealers involved.

\section{Conclusion}
\label{sec:conclusion}

At first the microstructure of the OTC credit index market seems different from that of equity and futures markets, as it is centered around dealers, without a central order book. However, we find that from the point of view of order flow and price impact, the differences are only quantitative. Client orders are much less split, and as a consequence, the impact of an isolated order, as expressed by $G$, reaches a permanent plateau. The numerical value of impact, after correcting for the imperfect identification of order signs, is the same order of magnitude as the bid-ask spreads in our daily trading experience. This is in line with what is expected based on theoretical arguments about the break-even costs of market making \cite{wyart2008relation}. The propagator takes $5-10$ trades, in real time more than $15$ minutes, to reach its final level. Because the market is very fragmented, it takes this long for the effect of a trade to become fully incorporated in the price.

Qualitatively the behavior is in line with what was observed for other, more frequently studied products. This finding gives further support to the argument that price impact is a universal phenomenon, and it behaves similarly in classical markets and more ``exotic" ones such as Bitcoin \cite{donier2015million}, options \cite{toth2016square} and now credit.

\section*{Acknowledgments}
The authors thank Panos Aliferis, Iacopo Mastromatteo and Bence T\'oth for their ideas and critical input.

\bibliography{bibs}{}

\end{document}